\documentclass[fleqn,usenatbib]{mnras}

\usepackage{newtxtext,newtxmath}

\usepackage[T1]{fontenc}

\DeclareRobustCommand{\VAN}[3]{#2}
\let\VANthebibliography\thebibliography
\def\thebibliography{\DeclareRobustCommand{\VAN}[3]{##3}\VANthebibliography}

\usepackage{graphicx}	
\usepackage{amsmath}	
\usepackage{float}

\title[CO molecular gas-rich galaxy overdensities]{H$_2$ molecular gas absorption-selected systems trace CO molecular gas-rich galaxy overdensities}

\author[A. Klitsch et al.]{
Anne Klitsch,$^{1}$\thanks{E-mail: anne.klitsch@gmail.com}
C\'eline P\'eroux,$^{2,3}$ 
Martin~A.~Zwaan,$^{2}$ 
Annalisa De Cia,$^{4}$ 
C\'edric Ledoux,$^{5}$ \newauthor{
Sebastian Lopez$^{6}$}
\\
$^{1}$DARK, Niels Bohr Institute, University of Copenhagen, Jagdtvej 128, 2200 Copenhagen, Denmark\\
$^{2}$European Southern Observatory, Karl-Schwarzschildstrasse 2, D-85748 Garching bei M\"unchen, Germany\\
$^{3}$Aix Marseille Univ, CNRS, CNES, LAM, (Laboratoire d'Astrophysique de Marseille), UMR 7326, 13388, Marseille, France\\
$^{4}$Department of Astronomy, University of Geneva, Chemin Pegasi 51, 1290 Versoix, Switzerland\\
$^5$European Southern Observatory, Alonso de C\'ordova 3107, Casilla 19001, Vitacura, Santiago, Chile\\
$^{6}$Departamento de Astronom\'ia, Universidad de Chile, Casilla 36-D, Santiago, Chile
}

\date{Accepted 2021 June 5. Received 2021 June 4; in original form 2021 January 20.}

\pubyear{2021}

\begin{document}
\label{firstpage}
\pagerange{\pageref{firstpage}--\pageref{lastpage}}
\maketitle

\begin{abstract}
Absorption-selected galaxies offer an effective way to study low-mass galaxies at high redshift. 
However, the physical properties of the underlying galaxy population remains uncertain. 
In particular, the multiphase circum-galactic medium is thought to hold key information on gas flows into and out of galaxies that are vital for galaxy evolution models. 
Here we present ALMA observations of CO molecular gas in host galaxies of H$_2$-bearing absorbers. 
In our sample of six absorbers we detect molecular gas-rich galaxies in five absorber fields although we did not target high-metallicity ($>50$ per cent solar) systems for which previous studies reported the highest detection rate.
Surprisingly, we find that the majority of the absorbers are associated with multiple galaxies rather than single haloes. 
Together with the large impact parameters these results suggest that the H$_2$-bearing gas seen in absorption is not part of an extended disk, but resides in dense gas pockets in the circum-galactic and intra-group medium.
\end{abstract}




\section{Introduction}

Conventional surveys identify galaxies based on the continuum or line emission and are therefore preferentially select the most luminous part of the galaxy populations. 
These are high stellar mass and star-formation rate galaxies. 
An alternative galaxy selection technique is based on the absorption signature that the gas content imprints in the spectra of otherwise unrelated background quasars \citep[damped Ly$\alpha$ absorbers with a column density of HI > $~2\times 10^{20}$~cm$^{-2}$][]{Wolfe2005}.
The absorption traces the neutral HI gas in galaxies. 
Damped Ly$\alpha$ absorbers trace the majority of HI in the Universe, as well as the majority of metals at high redshift \citep{Peroux2020a}. 
These absorption-selected galaxies include the selection of lower-mass galaxies that lie at or below the main sequence for star-forming galaxies at their respective redshifts \citep[e.g.][]{Kanekar2018, Rhodin2018}.

Therefore, a good understanding of the underlying galaxy population traced by absorption-selection is required to draw firm conclusions on the evolution of the whole galaxy population. 
It has been suggested that absorption-selected galaxies at intermediate redshift ($z \sim 0.5$) are extremely molecular gas-rich and have a low star formation efficiency compared to emission-selected galaxies at low ($z \sim 0$) and high ($z \sim 2$) redshift \citep{Kanekar2018,Kanekar2020}. 
The authors of these studies selected high metallicity absorbers and report CO detection rates of 40--70~per cent.
In another sample of six absorption-selected galaxies observed with MUSE and ALMA, the MUSE-ALMA Halos survey, \citet{Szakacs2021} report a CO detection rate of 1/3.
However, in total only small samples of molecular gas mass measurements at low, intermediate and high redshifts are available. 

At the same time the advent of large field-of-view integral field spectrographs in the optical such as the Multi Unit Spectroscopic Explorer (MUSE) on the VLT and the Keck Cosmic Web Imager (KCWI) on the Keck~II telescope enabled efficient detection of absorber host galaxies with star-formation rate limits of $\sim 0.1 \rm{M_{\odot} yr^{-1}}$. 
These surveys have revealed an increasing number of absorbers associated to groups of galaxies rather than individual systems \citep[e.g.][]{Bielby2017, Peroux2017, Klitsch2018, Peroux2019, Hamanowicz2019, Schroetter2020, Lofthouse2020}.
Recently, a group of nine galaxies associated with an H$_2$ absorber within 600 proper kpc tracing a galaxy overdensity was presented by \citet{Boettcher2020a}. 

A growing sample of molecular gas observations in absorption-selected systems is now available \citep[e.g.][]{Neeleman2016, Neeleman2018, Moller2018, Klitsch2018, Klitsch2019, Kanekar2018,Kanekar2020, Fynbo2018, Szakacs2021}
Only in two absorption-selected systems more than one molecular gas-rich galaxy have been reported in emission \citep{Klitsch2019, Peroux2019}. 
However, a systematic study in CO emission observed with ALMA similar to those for ionised gas carried out in the optical with MUSE is still missing. 

The goal of this study is to follow-up Ly$\alpha$ absorbers with previously identified H$_2$ absorption identified from UV absorption lines.
The presence of H$_2$-bearing gas in absorption potentially suggests a large reservoir of molecular gas also observable in CO emission. 
We search for molecular gas emission from the absorber host galaxy(s) and aim to identify the origin of the absorbing gas.

This paper is organised as follows: 
In Section 2 we describe the target selection, observations and data reduction, the detected galaxies and the relations between the absorbing gas and the host galaxies are presented in Section 3 and in Section 4 we give a summary of this work and list our conclusions.

Throughout the paper we adopt a $\Lambda$CDM cosmological model with $H_0 = 70 {\rm \; km \; s^{-1} \; Mpc^{-1}}$, $\Omega_{\rm m} = 0.3$, and $\Omega_{\Lambda}= 0.7$.

\begin{table*}
	\centering
	\caption{Physical properties of quasar absorbers targeted in this sample.}
	\label{tab:AbsProp}
	\begin{tabular}{lcccccccccc} 
		\hline
		Field & Alternative & $z_{\rm QSO}$ & $z_{\rm abs}$ & log(N(HI)/cm$^{-2}$) & log(N(H$_2$)/cm$^{-2}$) & log$Z$/$Z_{\odot}$ & $\Theta$\\
		 & Name & & & & & & [kpc] \\
		\hline
QSO B0120-28 & & 0.434 & 0.18561 & $20.50 \pm 0.10^d$ & $20.00 \pm 0.10^d$ & $-1.19 \pm 0.21^d$ &  $70^d$\\
J1342-0053 & LBQS 1340--0038 & 0.326 & 0.22711 & $19.0^{+0.5}_{-0.8}$$^e$ & $14.63 \pm 0.06^a$ & $-0.40^{+0.40}_{-0.10}$$^k$ & $35^e$\\
Q2131--1207 & PKS 2128--123 & 0.500 & 0.4298 & $19.5\pm 0.15^j$ & $16.36\pm 0.08^j$ & $> -0.96^b$ & 12, 48$^f$\\
Q1241+176  & QSO J1244+1721 & 1.273 & 0.55048 & $>19.00^i$ & $15.81 \pm 0.17^a$ & $< 0.18^a$ & 21$^g$\\
Q0107-0232 & LBQS 0107-0232 & 0.728 & 0.55733 & $19.50 \pm 0.20^h$ & $17.27 \pm 0.30^h$ & $-0.72 \pm 0.32^h$ & 10, 11$^h$\\
HE 0515-4414  & QSO B0515-4414 & 1.713 & 1.15 & $19.88 \pm 0.05^c$ & $16.94^{+0.23}_{-0.41}$$^c$ & $-0.38 \pm 0.04^l$\\
		\hline
	\end{tabular}
	\flushleft{
	\textbf{Notes:} $\Theta$ denotes the impact parameter between the quasar sight-line and a previously identified absorber host galaxy. \textbf{References:} $^a$~Average metallicities ([S/H] or
[Si/H]) from \citet{Muzahid2015} without ionization correction, $^b$~\citet{Som2015}, $^c$~\citet{Reimers2003}, $^d$~\citet{Oliveira2014}, $^e$~\citet{Werk2013}, $^f$~\citet{Peroux2017}, $^g$~\citet{Steidel1994}, $^h$~\citet{Crighton2013}, $^i$ \citet{Rao2006}, $^j$ \citet{Muzahid2016}, $^k$ \citet{Werk2014}, $^l$\citet{Quast2008}
}
\end{table*}

\section{Observations and Data Reduction} \label{Section2}

\subsection{Target selection}
The absorbers were selected to have an H$_2$ molecular gas absorption detection at UV wavelength. 
Additionally, we selected absorbers observable with ALMA and  at $z \lesssim 1$ to limit the required observing time. 
The neutral gas column densities of these absorbers are in the range of log(N(HI)/${\rm cm^{-2}}$) = $19.00 - 20.50$ and the molecular fractions cover a range of log~$f_{\rm mol} = -4$ to $-0.4$. 
We find that although it was not a selection parameter, some absorbers in our sample have a high metallicity of above 30 per cent solar, but we also include low metallicity systems with only $\sim 10$ per cent of solar metallicity. We note that some of these metallicity estimates do not include ionisation corrections and could therefore be somewhat higher.
This distribution is expected for a sample of H$_2$ absorbers \citep{Bolmer2019}. 
The absorber properties including optically identified host galaxies are listed in Table~\ref{tab:AbsProp}.

\subsection{ALMA observations}

The six target fields are listed in Table \ref{tab:AbsProp}. 
We observe the lowest possible CO transition for each absorber where the specific transition is determined by the absorber redshift ($z_{\rm abs}$). 
The lowest CO transition is observed to minimize uncertainties from the line luminosity conversion, needed to evaluate the total molecular gas mass.
The observations are carried out under program 2018.1.01575.S (PI Klitsch). 
All observations use four ALMA 1.875 GHz bands, with one band (in spectral-line mode of the correlator) covering the respective expected redshifted CO line frequency, and the other three bands (in continuum mode) placed at neighbouring frequencies to measure the continuum emission.
We start the data reduction with the pipeline calibrated uv-data sets, as delivered by ALMA. 
Additional data reduction steps are carried out with the Common Astronomy Software Applications (\textsc{CASA}) software package version 5.6 \citep{McMullin2007}. 
The imaging is done using the standard \textit{tclean} algorithm with a final field of view of 1.5 times the primary beam size. 
This results in an image diameter of $1.4\arcmin$ in ALMA Band 3 and $1.2\arcmin$ in ALMA Band 4.  
A `natural' weighting scheme is applied, which guarantees optimal sensitivity appropriate for a line detection experiment. 
The final beam sizes as well as the rms noise are listed in Table \ref{tab:ObsProp}. 
We use a channel width of 15 - 20 km s$^{-1}$. 

\begin{table}
	\centering
	\caption{Details of the ALMA observations.}
	\label{tab:ObsProp}
	\begin{tabular}{lcccccccccc} 
		\hline
Field & CO & ALMA & RMS & beam & beam \\
 & line & Band & [mJy/beam] & [\arcsec~$\times$~\arcsec] & [kpc~$\times$~kpc]\\
\hline
QSO J0120-28 & 1-0 & 3 & 0.62 & 1.0 x 0.8 & 3 x 3\\
Q1342-0053 & 1-0 & 3 & 0.33 & 0.9 x 0.8 & 3 x 3\\
PKS 2128-123   & 2-1 & 4 & 0.90 & 0.9 x 0.8 & 5 x 5\\
PG 1241+176 & 2-1 & 4 & 0.50 & 1.6 x 1.4 & 11 x 9\\
Q0107-0232 & 2-1 & 4 & 0.42 & 1.6 x 1.3 & 10 x 9\\
HE 05151-4414 & 3-2 & 4 & 0.34 & 1.5 x 1.2 & 12 x 10\\
		\hline
	\end{tabular}
\end{table}

\subsection{Archival MUSE observations} \label{Sec:DataMUSE}

One of the ALMA targets (HE 0515-4414) was observed with MUSE under the project ID: 1100.A-0528 (PI: Fumagalli).
The MUSE FoV covers a $1\arcmin \times 1\arcmin$ area, slightly {narrower than that of our ALMA observations. 
We have downloaded the pipeline-processed and flux calibrated data cube from the ESO Phase 3 archive. 
We use this data to search for optical counterparts for the CO detected galaxies.

\begin{table*}
	\centering
	\caption{Molecular gas physical properties of galaxies associated with the absorbers.}
	\label{tab:GalOverview}
	\begin{tabular}{lccccccccccc} 
		\hline
		Name & Flag & RA & Dec & $\Theta$ & $v_0$ & FWHM & $S_{\rm int}$ & $S_{\rm peak}$ & $L'_{\rm CO}\times 10 ^{9}$ & M$_{\rm mol}\times 10 ^{9}$\\
		& & & & [kpc] & [km s$^{-1}$] & [km s$^{-1}$] & [Jy km s$^{-1}$] & [Jy] &  [K km s$^{-1}$] &  [M$_{\odot}$]\\
		\hline
QSO J0120--28  & - & - & - & - & - & - & $<$ 0.94 & - & $<$ 1.6 & $<$ 6.8\\
QSO J1342--0053 G1 & H & 13h42m51.85s & -00d53m54.2s & 35 & -9 & 165 $\pm$ 8 & 0.59 $\pm$ 0.06 & 4.0 $\pm$ 0.7 & 1.5 $\pm$ 0.1 & 6.5 $\pm$ 0.6\\
Q2131--1207 G1 2-1 & - & - & - & 52 & - & - &  $<$ 1.57 & - & $<$ 3.8 & $<$ 8.2\\
Q2131--1207 G1 3-2$^a$ & H & & & 52 & & 184 $\pm$ 50 & 0.355 & & & 5.29\\
PG 1241+176 G1 & H & 12h44m09.59s & +17d20m58.1s & 120 & 620 & 500 $\pm$ 10 & 1.49 $\pm$ 0.19 & 7.0 $\pm$ 1.8 & 5.9 $\pm$ 0.8 & 12.8 $\pm$ 1.6\\
PG 1241+176 G2 & H & 12h44m10.00s & +17d20m58.1s & 86 & 460 & 300 $\pm$ 10 & 0.64 $\pm$ 0.09 & 3.0 $\pm$ 0.8 & 2.5 $\pm$ 0.3 & 5.5 $\pm$ 0.7\\
PG 1241+176 G3 & H & 12h44m10.64s & +17d20m55.6s & 60 & 100 & 360 $\pm$ 10 & 0.41 $\pm$ 0.06 & 2.2 $\pm$ 0.6 & 1.6 $\pm$ 0.2 & 3.5 $\pm$ 0.5\\
LBQS 0107--0232 G1 & L & 01h10m14.22s & -02d17m06.6s & 63 & -291 & 60 $\pm$ 10 & 0.13 $\pm$ 0.02 & 2.4 $\pm$ 0.6 & 0.5 $\pm$ 0.1 & 1.1 $\pm$ 0.2\\
LBQS 0107--0232 G2 & L & 01h10m15.56s & -02d17m12.4s & 144 & 449 & 240 $\pm$ 10 & 0.39 $\pm$ 0.07 & 4.7 $\pm$ 1.0 & 1.6 $\pm$ 0.3 & 3.4 $\pm$ 0.6\\
LBQS 0107--0232 G3 & L & 01h10m14.82s & -02d17m06.6s & 69 & -651 & 140 $\pm$ 10 & 0.17 $\pm$ 0.03 & 1.7 $\pm$ 0.5 & 0.7 $\pm$ 0.1 & 1.5 $\pm$ 0.2\\
HE 05151--4414 G1 & H & 05h17m08.51s & -44d10m51.4s & 82 & -188 & 600 $\pm$ 10 & 0.80 $\pm$ 0.06 & 2.6 $\pm$ 0.5 & 6.2 $\pm$ 0.5 & 7.2 $\pm$ 0.5\\

HE 05151--4414 G2 & L & 05h17m07.12s & -44d11m01.5s & 67 & -213 & 105 $\pm$ 8 & 0.08 $\pm$ 0.01 & 0.8 $\pm$ 0.3 & 0.6 $\pm$ 0.1 & 0.7 $\pm$ 0.1\\
		\hline
	\end{tabular}
	\flushleft
	\textit{Notes:} 
	The flag column shows the confidence (H: high, L: low) we assign to the detection. 
	For more information see Section~\ref{text:confidenceFlags}. 
	$Theta$ denotes the impact parameter measured between the QSO sight line and the galaxy position. 
	$v_0$ is the velocity difference between the H$_2$ velocity centroid and the CO emission of the galaxy.  
	The uncertainties in the molecular gas mass only include the flux measurement uncertainty, uncertainties in the conversion factors are discussed in Section \ref{text:molecularGasMass}. 
	Upper limits are calculated assuming a 5$\sigma$ detection threshold and a linewidth of 300~km~s$^{-1}$. 
	For QSO J0120-28 we use the sensitivity in the center of the image cube since no reliable counterpart has been reported. 
	For Q2131-1207 we give the upper limit at the position of the optical counterpart reported by \cite{Peroux2017} and \cite{Szakacs2021}. 
	$^a$ In addition, we list as Q2131-1207 G1 3-2 the CO(3-2) measurement reported by \citet{Szakacs2021}.
\end{table*}

\section{Results}\label{Section3}

We search for CO line emission in the ALMA image cubes before correcting for the primary beam response to ensure constant noise properties across the field of view. 
We search for line emission both by-eye and using the line finder software \textsc{SoFiA} and detect CO line emission in four out of six absorber fields. 
We find in total nine galaxies. 
Three out of four absorbers have more than one galaxy in close proximity. 
In addition to the detections from this work we use the results for Q2131-1207 presented by \citet{Szakacs2021}. 
The properties of our detections are listed in table \ref{tab:GalOverview}.

The velocity profiles of the detected galaxies have a FWHM of 60---500\,km\,s$^{-1}$. 
Previously reported CO detections of absorption selected galaxies have a FWHM of $> 300$\,km\,s$^{-1}$ \citep[e.g.][]{Kanekar2018, Kanekar2020, Klitsch2018, Klitsch2019, Moller2018, Peroux2019}. 
While the narrow CO emission lines could trace low mass galaxies we also note that the detection is only significant over few channels in the ALMA data.
Therefore, we flag galaxies with a FWHM $< 300$\,km\,s$^{-1}$ as low significance detections unless the detection is confirmed by independent observations. 
These are still included in the further analysis to avoid a bias against low mass galaxies. 
When studying trends and drawing general conclusions we regard both, the high significance sample only and the full sample.
\label{text:confidenceFlags}

\subsection{Notes on the individual quasar fields}

Emission line maps and extracted spectra of the detections are shown in Fig.~\ref{fig:EmLineMaps} and \ref{fig:VelocityProfiles}.
The relation between the CO detection and the absorber in velocity space is illustrated in Fig.~\ref{fig:AbsEmsOverlay}. 
The CO spectra are obtained from the primary beam corrected image cubes. 
Since many sources are resolved we integrate the spectra over a region defined by the 3$\sigma$ contour in the integrated flux map. 
We do not detect significant continuum emission from any galaxy detected in CO. 
A detailed analysis of the different fields is presented in the following.

\begin{figure*}
    \centering
    \includegraphics[width =0.24\linewidth]{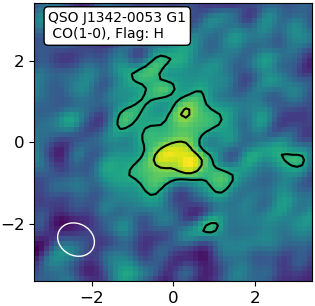}
    \includegraphics[width =0.24\linewidth]{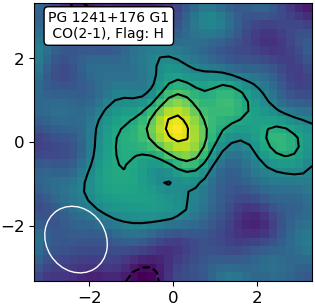}
    \includegraphics[width =0.24\linewidth]{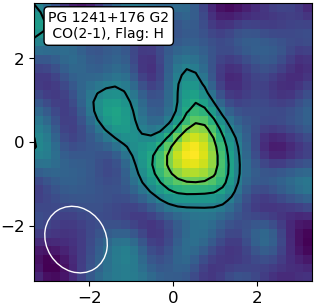}
    \includegraphics[width =0.24\linewidth]{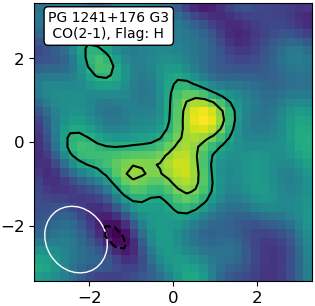}
    \includegraphics[width =0.24\linewidth]{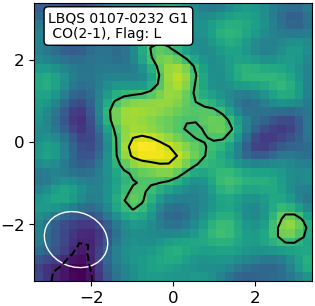}
    \includegraphics[width =0.24\linewidth]{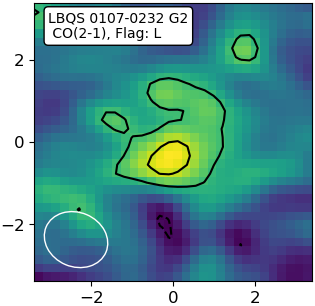}
    \includegraphics[width =0.24\linewidth]{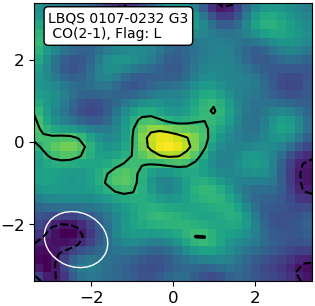}\\
    \includegraphics[width = 0.24\linewidth]{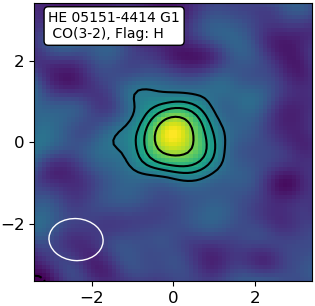}
    \includegraphics[width = 0.24\linewidth]{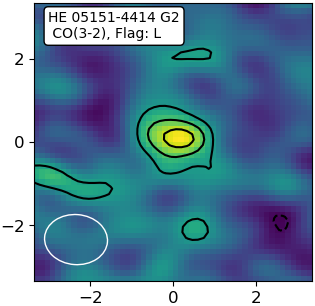}
    \caption{CO emission line maps for all absorption-selected galaxies in this sample. The maps show a $6 \times 6 \arcsec$ wide field. 
    Contours mark the 3, 5, 7, 10 sigma level in the integrated intensity plot, negative contours are plotted as dashed lines, the white ellipse shows the ALMA synthesized beam, the spectrum is extracted from the region where the integrated flux map is above 3 sigma. We also report the emission line and the confidence flag in the plot labels. The confidence flags (H: high, L: low) are discussed in Section~\ref{text:confidenceFlags}.  We report nine CO detections of host galaxies in close proximity to four absorbers with three absorbers showing significant galaxy overdensities.}
    \label{fig:EmLineMaps}
\end{figure*}

\begin{figure*}
    \centering
    \includegraphics[scale = 0.6]{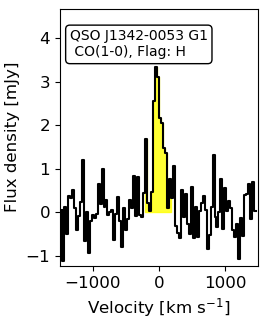}
    \includegraphics[scale = 0.6]{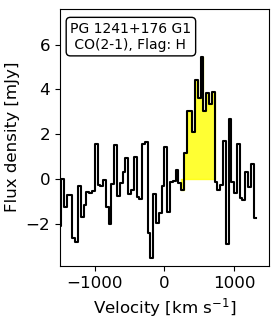}
    \includegraphics[scale = 0.6]{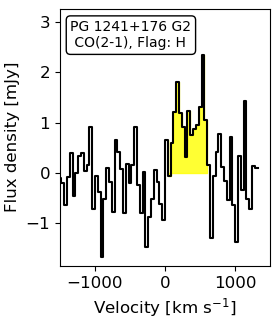}
    \includegraphics[scale = 0.6]{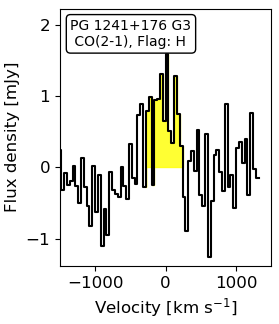}
    \includegraphics[scale = 0.6]{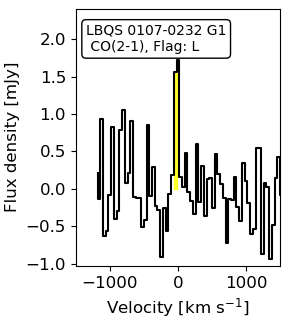}
    \includegraphics[scale = 0.6]{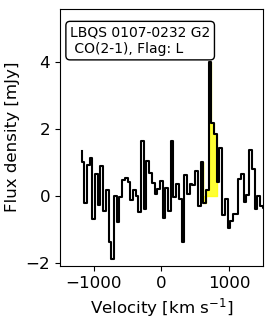}
    \includegraphics[scale = 0.6]{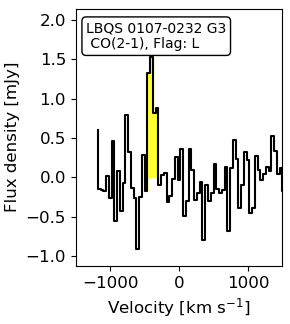}\\
    \includegraphics[scale = 0.6]{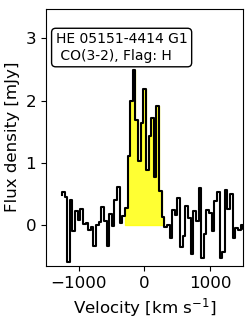}
    \includegraphics[scale = 0.6]{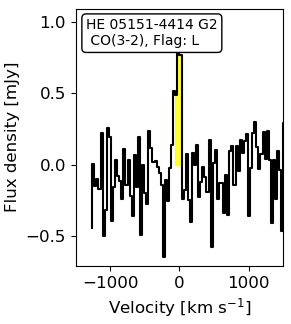}
    \caption{
    CO emission line spectra for all absorption-selected galaxies in this sample. 
    The spectra are extracted from the region where the integrated intensity maps in Fig.~\ref{fig:EmLineMaps} are above 3 sigma, spectra are PB corrected and re-binned for display. The zero velocity is set to the systemic velocity of the galaxy with the smallest impact parameter. We also report the emission line and the confidence flag in the plot labels. The confidence flags (H: high, L: low) are discussed in Section~\ref{text:confidenceFlags}.}
    \label{fig:VelocityProfiles}
\end{figure*}

\paragraph*{B0120-28 ($z_{\rm abs} = 0.18562$)}

We do not detect a host galaxy associated with the absorber in this field. 
\citet{Oliveira2014} report an absorber host galaxy at an impact parameter of 70~kpc based on imaging and spectroscopic data. 
However, the full details of this galaxy remain unpublished. 
We have therefore calculated an upper limit on the molecular gas mass based on the sensitivity limit of our CO(1--0) emission line observations.

\paragraph*{J1342-0053 ($z_{\rm abs} = 0.22711$)}

We report a detection of CO(1--0) emission coincident with a previously optically identified absorber host galaxy \citep{Werk2013}.
The authors report a SFR of 6.04M$_{\sun}$ yr$^{-1}$, a stellar mass of $10^{10.93}$M$_{\sun}$, a luminosity of 1.08$L^{\star}$, and a super solar metallicity of 12 + log(O/H) = 9.05.

\paragraph*{Q2131-1207 ($z_{\rm abs} = 0.4298$)}

We do not detect CO(2--1) emission in the field of Q2131-1207 at the sensitivity limit of our observations. 
This field has been observed with VLT/MUSE presented by \citet{Peroux2017}. 
We carefully examine the positions of the galaxies presented by the authors of this work, however, we also do not detect any low confidence emission from these galaxies. 
\citet{Szakacs2021} have followed-up this field in CO(3--2) to a higher SNR. 
They report CO emission from one associated galaxy (SFR$ = 2.0 \;\rm{M_{\odot}}\; \rm{yr}^{-1}$, L$ = 0.6\;\rm{L}_{\star}$). 
We calculate an upper limit of the CO(2-1) emission at the position of this galaxy which is consistent with the detection by \citet{Szakacs2021}
\citep[assuming a Milky Way type CO line ratio $L'_{\rm CO(3-2)}/L'_{\rm CO(2-1)} = r_{32} = 0.54$; ][]{Bolatto2013}.
We include this field in the further analysis using the CO(3--2) detection.

\paragraph*{Q1241+176 ($z_{\rm abs} = 0.55048$)}

In this field, we detect three galaxies at the absorber redshift. 
The galaxies lie at impact parameters of 60 -- 120~kpc. 
In addition, \citet{Steidel1994} have identified a host galaxy at an impact parameter of $21.2 \pm 0.3$kpc.
However, the exact details of this detection are not reported.
We carefully search for CO emission at this impact parameter, but do not report a host galaxy.

\paragraph*{Q0107-0232 ($z_{\rm abs} = 0.55733$)}

\citet{Crighton2013} have identified two galaxies within $<2\arcsec$ of the quasar sight-line in K-band imaging. 
If these galaxies were at the absorber redshift this would translate to impact parameters of 10 kpc and 11 kpc. 
We do not find CO emission from the position of these galaxies.
However, we note that there is no redshift information available for these galaxies and these could be a chance alignment. 
We therefore exclude these galaxies as the absorber hosts. 
Instead, we detect three galaxies at impact parameters of 60 -- 140~kpc. 
Given the narrow FWHM of these CO emission lines we label those as low confidence detections. 
In the further analysis these galaxies are excluded when drawing general conclusions.

\paragraph*{HE 0515--414 ($z_{\rm abs} = 1.15$)}

We have identified two galaxies at the absorber redshift (one high and one low confidence) with impact parameters of 67 and 85\,kpc. 

This field was also observed using WFC3/G141 HST GRISM spectroscopy (PI: Bielby, ID: 14594). 
They have detected 9 galaxies at $z = 1.14-1.16$ within a field-of-view of 6' out of which one is G1 that we detect in CO (Bielby private comm.).
Other galaxies identified in these data are outside the field-of-view of our ALMA observations.

We have obtained additional archival MUSE observations described in Section~\ref{Sec:DataMUSE}. 
We search for counterparts of the CO detected galaxies. 
At the absorber redshift the [OII] line is covered by the MUSE spectral range. 
We detect [OII] emission from G1, but not from G2. 
The [OII] emission line profile and emission line map are shown in Fig.~\ref{fig:OIIemission}.
Following the description by \citet{Kennicutt1998} we measure a non dust-corrected SFR of 9.4~M$_{\odot}$\,yr$^{-1}$ which represents a lower limit on the true SFR.
This translates to an upper limit for the molecular gas depletion time of 0.8~Gyr. 
This is in contrast to the long gas depletion times for absorption selected galaxies at intermediate redshifts reported by \citet{Kanekar2018, Peroux2019} and \citet{Szakacs2021}.

Additionally, we use the [OII] and CO observations of G1 to study the ionised and molecular gas kinematics. 
The line of sight velocity maps are shown in Fig.~\ref{Fig:VelMaps}. 
The CO line emission is only marginally resolved and the [OII] emission is not significant enough to allow for a detailed analysis using state-of-the-art forward modelling techniques.
However, a qualitative analysis shows that the molecular and ionized gas kinematics are coupled. 
Both show rotation with similar orientation and maximum velocity. 

\subsection{Molecular gas masses} \label{text:molecularGasMass}

We convert the CO emission line luminosities in the observed transitions to the CO(1-0) line luminosity using a Milky Way type line ratio as reported by \citet{Carilli2013}. 
As we have no further information on the CO line excitation this assumption yields an upper limit of the molecular gas mass. 
We stress that previous studies of the CO spectral line energy distribution of absorption-selected galaxies have in some cases shown more excited ISM conditions than that of the Galaxy \citep{Klitsch2019}.
The CO(1--0) line fluxes could therefore be overestimated by a factor of 2 -- 4.
We have used a CO-to-H$_2$ conversion factor of $\alpha_{\rm CO} = 4.36 \rm{M}_{\sun}\: (\rm{K \: km\: s^{-1} pc^2})^{-1}$ that is applicable for galaxies of solar metallicity that do not undergo a starburst \citep[including a factor of 1.36 to account for the presence of helium ][]{Bolatto2013}.
We caution, however, that the true molecular gas mass might be overestimated by up to a factor of six compared to $\alpha_{\rm CO} = 0.8 \rm{M}_{\sun}\: (\rm{K \: km\: s^{-1} pc^2})^{-1}$, appropriate for starburst environments. 
In case of a systematically too high conversion factor chosen for all galaxies the relative trends remain the same. If, however, only some galaxies in this sample have a Milky Way type CO line ratio and $\alpha_{\rm CO}$ while for others a starburst type conversion is more appropriate this could introduce also relative shifts.

\section{Discussion}

\begin{figure}
\centering
\includegraphics[width = 0.9\linewidth]{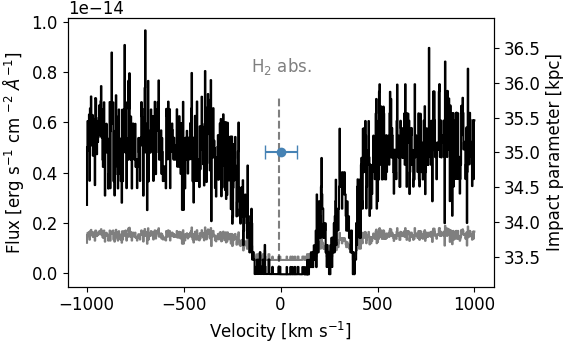}
\includegraphics[width =  0.9\linewidth]{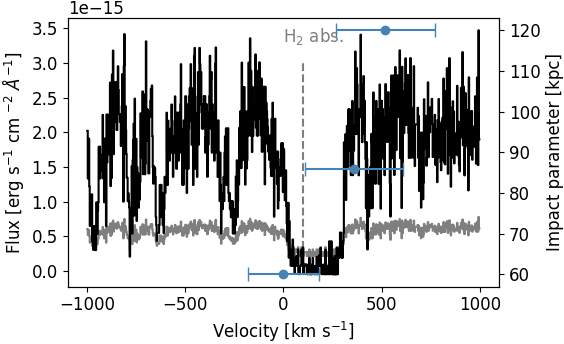}
\includegraphics[width = 0.9 \linewidth]{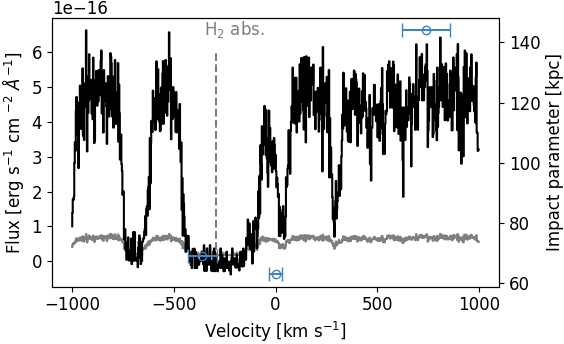}
\includegraphics[width = 0.9 \linewidth]{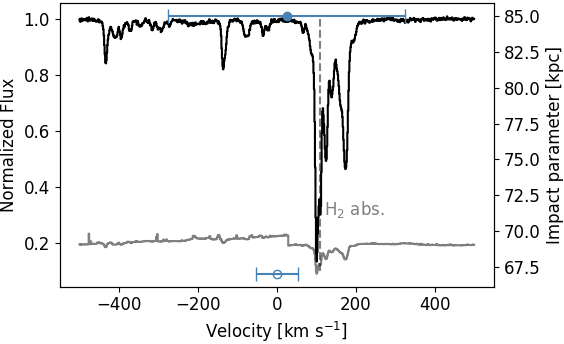}
\caption{Comparison between absorption profiles and the galaxy detections in CO. The absorption lines shown are QSO1342: Ly$\alpha$ (HST COS), PG1241: Ly$\delta$ (HST COS), LBQS0107: Ly$\delta$ (HST COS), HE 0515: MgI 2852\AA (UVES). The zero velocity refers to the systemic velocity of the host galaxy with the smallest impact parameter. Blue points mark the velocity offset of the CO detections, the errorbar width marks the FWHM, the vertical position marks the impact parameter from the quasar (see right axis). Filled and open symbols represent the high and low significance flags described in Section \ref{Section3}. 
The grey vertical dashed line marks the centroid of the H$_2$ absorption. Different absorption components around the main Ly transition are interlopers from the dense Ly$\alpha$ forest. The MgI absorption for HE0515 shows several velocity components spreading from $-100$ -- 250~km\;s$^{-1}$ that all belong to the same absorber.\label{fig:AbsEmsOverlay}}
\end{figure}

\begin{figure}
\includegraphics[width = \linewidth]{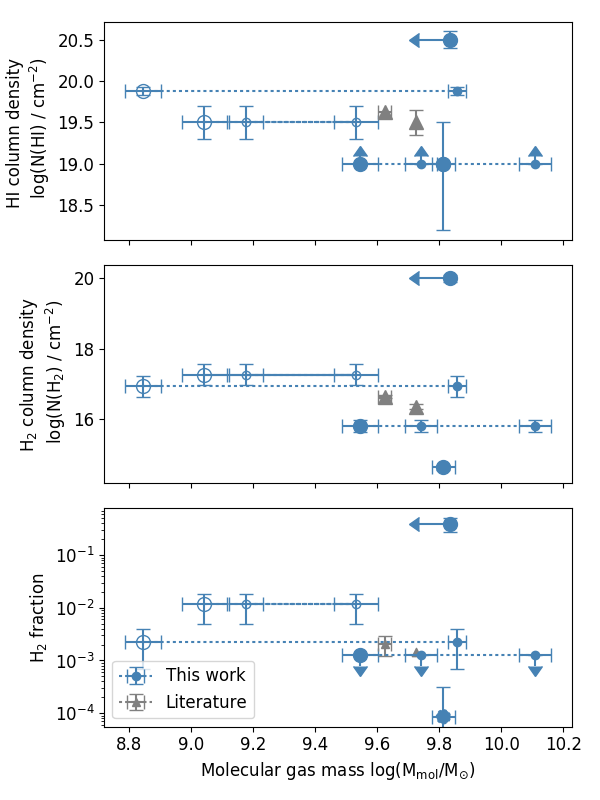}
\caption{
HI and H$_2$ absorber column density and molecular gas fraction as a function of molecular gas mass of the host galaxies (top to bottom). 
Larger markers indicate the galaxy with the smallest impact parameter. 
Filled and open symbols represent the high and low significance flags described in Section \ref{Section3}. 
Grey markers represent literature measurements from \citep{Neeleman2016, Szakacs2021}. 
We report no clear trend of absorption properties with the molecular gas mass of the host galaxies in both the full and the high significance sample. This suggests that H$_2$ absorption is tracing pockets of H$_2$-bearing gas in the outskirts of the CGM rather than an extended molecular disk. \label{fig:relationsMolecularGasMass}}
\end{figure}

\begin{figure}
\includegraphics[width = \linewidth]{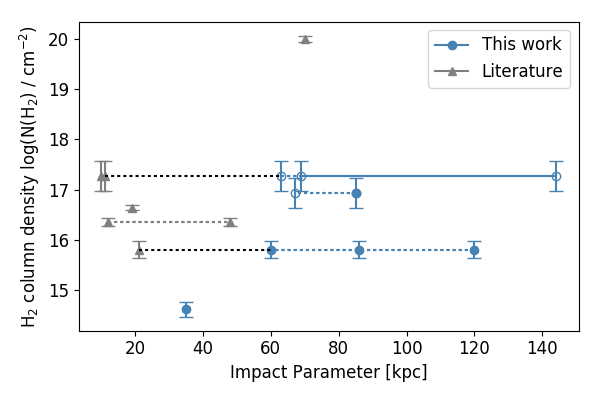}
\caption{
H$_2$ column density as a function of impact parameter for the detected galaxies in this work (blue, filled - high significance, open - low significance) and galaxies presented in the literature \citep{Neeleman2016, Oliveira2014, Crighton2013, Szakacs2021}.
The lack of a correlation between H$_2$ molecular gas column density and the impact parameter in both samples (full and high significance) suggests that the absorption is not tracing extended molecular gas distributions. \label{fig:H2colDensVSimpactParam}}
\end{figure}

\subsection{Detection rate}

In this study we have selected the targets based on previously identified H$_2$-bearing gas detected in absorption instead of a metallicity based selection that has sometimes been used in previous such surveys. 
We find that the absorbers cover a range of 1/10$^{\rm th}$ to solar metallicity.
We list the molecular gas column density as well as the metallicity of the absorbers in Table~\ref{tab:AbsProp}. 
We find molecular gas-rich galaxies in five out of six fields and we detect multiple galaxies in three absorber fields. 
It is known from archival MUSE observations that in a fourth field (Q2131-1207) an overdensity of four galaxies is detected \citep[][]{Peroux2017}. 
This success rate of $>80$ per cent is remarkable for a study of molecular gas in absorption-selected systems.
Similar studies at intermediate and high redshift focussing on high metallicity absorbers achieved a detection rate of $\sim 40 - 70$ per cent with a clear relation to the absorber metallicity \citep{Kanekar2018, Kanekar2020, Szakacs2021}. 
We attribute the higher detection rate of CO emission to the presence of the H2-bearing clouds in absorption that indicate the presence of particularly molecular gas-rich systems.

\subsection{Absorption associated to galaxy overdensities rather than single galaxies}

A remarkable finding of this work is the detection of multiple galaxies associated to one absorber in most fields \citep[three out of four fields presented here and in Q2131-1207 presented by][]{Peroux2017, Szakacs2021}. 
For the field of HE0515-4414 we report an even more extended structure of galaxies including 10 galaxies within $\sim 3$~Mpc.
A similar finding was reported by \citet{Hamanowicz2019} who present a sample of absorption-selected galaxies with follow-up VLT/MUSE IFU observations.

We test if the observed multiple galaxies are indeed overdensities by calculating the expected number of galaxies within the co-moving volume traced by our observations.
We use the CO luminosity function reported by \citet{Fletcher2020} and integrate with a lower limit of $10^9\; \rm{K \;km \;s^{-1}}$ which is equivalent to the faintest detections in our sample. 
We find that at both, the low and high redshift end of our targets we expect fewer than one galaxy within the volume probed by the ALMA Band 3 and 4 observations. 
Here we consider a velocity range of $\pm 1000$~km~s$^{-1}$from the absorber redshift. 
This suggests that the multiple galaxies detected in each field represent overdensities in the galaxy distribution at the absorber redshift.

\subsection{Molecular gas mass relation to absorber properties}

Although the sample size is limited and the connection between absorber and host galaxies might be complex due to identification of multiple host galaxies, we test if we can find a relation between the molecular gas mass of the absorption-selected galaxies and the absorber properties.
In Fig.~\ref{fig:relationsMolecularGasMass} we show the H{\sc i} column density, the H$_2$ column density and the molecular gas fraction as a function of the molecular gas mass of the host galaxies.
We do not report a significant correlation in any of these parameter spaces for both, the full sample and the high significance detections only.
This is probably expected as the connection between the absorber and the host galaxy in such complex systems is not straightforward. 
A similar non-correlation was identified in optical studies finding galaxy overdensities related to absorbers in a similar redshift range \citep{Hamanowicz2019}.

\subsection{Do molecular gas absorbers probe disks of galaxies or pockets of dense gas in the CGM?}

To test whether the H$_2$-bearing gas traces disks of galaxies or pockets of dense gas in the circum-galactic medium (CGM) we analyse the distance between the absorber and the host galaxies in both projected distance on sky (impact parameter) and distance in velocity space to probe the full extent of the three dimensional structure. 
In Fig.~\ref{fig:AbsEmsOverlay} we show the absorption profiles in relation with emission line width of the CO detected galaxies in terms of both sky position (right y-axis) and velocity space (x-axis).
The lack of comparison spectra of strong metal absorption lines often hampers a direct comparison. 
For some sight lines the only available lines are embedded in the Ly$\alpha$ forest making a comparison between the absorption components and CO emission lines challenging. 

We find large impact parameters of 35 -- 140~kpc suggesting that the molecular absorption does not probe the extended molecular disk of a single galaxy. 
In addition, we find a good agreement between the absorption and CO emission in velocity space.
This suggests an association between the neutral and H$_2$-bearing gas at the position of the quasar sight line and the molecular gas in the host galaxies.

Additionally, we test for a relation between the H$_2$ column density and the impact parameter (see Fig.~\ref{fig:H2colDensVSimpactParam}), that would be expected if the absorption is tracing the disk of the host galaxy.
We do not identify a clear trend between these two quantities adding further evidence that the absorption is tracing pockets of H$_2$-bearing gas rather then the disk of a molecular gas-rich galaxy.
Comparing the detection rates of our work with previous studies focussing on high metallicity we find that first with a success rate of $> 80$~per cent our detection rate is significantly higher and second we find overdensities of galaxies rather then single galaxies.
This suggests that there is possibly more diffuse gas around overdensities of galaxies which provides the appropriate conditions for molecular gas to form in-situ or survive the outflow from a galaxy traced in absorption.

Galaxy interactions are numerous in overdense environments. 
Interactions can produce tidal streams as well as promote starburst or AGN driven outflows.  
The H$_2$ molecular gas could therefore be part of self-shielded, tidally stripped or ejected disk material. 
This scenario is supported by the observations of molecular gas in high velocity clouds in the Milky Way \citep{Richter2001, Putman2012}.
Furthermore, the molecular gas could be transported out to large distances by outflows \citep{Richings2018}. 

We note that we cannot exclude the presence of low molecular gas mass galaxies below our detection limit at small impact parameters. 
Ideal follow-up observations to settle this question would be optical IFU observations to detect all galaxies in these fields down to low star formation rates.

\section{Conclusions}

In this paper we present a systematic search for molecular gas traced by CO emission in H$_2$ absorption selected systems. 
We report the detection of CO emission from galaxies in close proximity to absorbers in five out of six quasar fields \citep[including the deeper observations presented by][]{Szakacs2021}. 
Our main conclusions can be summarised as follows:

\begin{itemize}
\item Our detection rate of $> 80$~per cent is higher than that reported in previous studies \citep{Kanekar2018, Kanekar2020} suggesting that the H$_2$-bearing gas absorbers trace CO molecular gas-rich galaxies.
\item Interestingly, the only system without a CO detection has the lowest metallicity. However, the CO detected galaxies are related to absorbers with a wide range of absorption metallicities.
\item The H$_2$-bearing gas in absorption in all except two sight-lines traces galaxy overdensities. In the case of HE0515-4414 this is also supported by GRISM spectroscopy tracing scales of 3\,Mpc. In the case of Q2131-1207 only one galaxy is detected in CO, but additional MUSE observations reveal three additional [OIII] emitting galaxies connected to the absorber. This adds further evidence that absorption selection is efficient at tracing overdense environments.
\item In the case of HE0515-4414 G1, we gather first insights in the molecular and ionised gas kinematics traced by ALMA and MUSE observations, respectively. We find that both velocity fields qualitatively agree in orientation and maximum velocity. Deeper high resolution observations are necessary for a complete morpho-kinematic modelling.
\item We detect galaxies at impact parameters of 35 -- 144 kpc. We conclude that the H$_2$-bearing absorbers do not trace the inner disk of a galaxy, but pockets of diffuse molecular gas in the circum-galactic and intra-group medium. This is also supported by the lack of a clear correlation between the absorber column density or molecular gas fraction with impact parameter or host galaxy molecular gas mass. 
\end{itemize}

\section*{Acknowledgements}

The authors thank Palle M\o ller for important discussions related to this work and Roland Szakacs for sharing preliminary results.
A.K.~thanks Aleksandra Hamanowicz for important discussions related to this work.
A.K.~gratefully acknowledges support from the Independent Research Fund Denmark via grant number DFF 8021-00130.
A.D.C.~acknowledges support by the Swiss National Science Foundation under grant 185692. 
S.L.~was funded by FONDECYT grant number 1191232.
This paper makes use of the following ALMA data: ADS/JAO.ALMA\#2018.1.01575.S. ALMA is a partnership of ESO (representing its member states), NSF (USA) and NINS (Japan), together with NRC (Canada), MOST and ASIAA (Taiwan), and KASI (Republic of Korea), in cooperation with the Republic of Chile. The Joint ALMA Observatory is operated by ESO, AUI/NRAO and NAOJ. In addition, publications from NA authors must include the standard NRAO acknowledgement: The National Radio Astronomy Observatory is a facility of the National Science Foundation operated under cooperative agreement by Associated Universities, Inc. 
This research made use of Astropy\footnote{\url{http://www.astropy.org}} a community-developed core Python package for Astronomy\citep{astropy2013, astropy2018}.
This research has made use of NASA's Astrophysics Data System.

\section*{Data Availability}

The data underlying this article are available in the respective observatories online archives. 
Used ALMA data is available through the ALMA Science Archive (\url{https://almascience.eso.org/asax/}),
VLT/MUSE data is available through the ESO SciencenArchive Facility (\url{http://archive.eso.org/cms.html}) 
and HST data through the Hubble Legacy Archive (\url{https://hla.stsci.edu}).
Project IDs are given in Section~\ref{Section2} and \ref{Section3}.




\bibliographystyle{mnras}
\bibliography{MyLibraryNoLink} 




\appendix

\section{Line of sight velocity maps of the ionised and molecular gas in HE0515-4414 G1}

\begin{figure*}
\includegraphics[width = 0.48\linewidth]{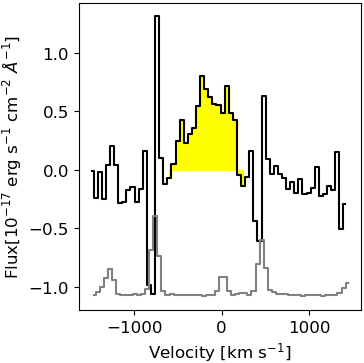}\hfill
\includegraphics[width = 0.48\linewidth]{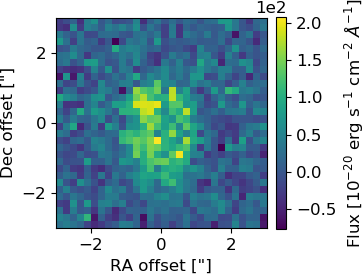}
\caption{Ionized gas observations of HE 0515-4414 Galaxy 1 ($z = 1.15$) traced by [OII] emission observed using MUSE. \textit{Left:} spectrum of the [OII] emission line, where the zero velocity marks the position of the absorber. \textit{Right:} integrated intensity map of the [OII] emission line integrated over the yellow region marked in the spectrum.\label{fig:OIIemission}}
\end{figure*}

\begin{figure*}
\includegraphics[width = 0.48\linewidth]{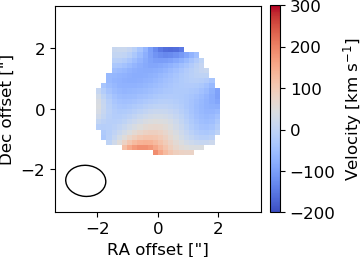}\hfill
\includegraphics[width = 0.48\linewidth]{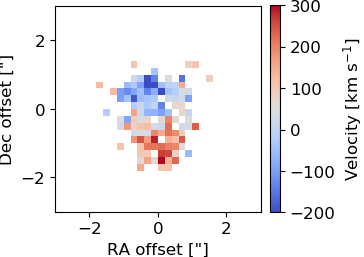}
\caption{Line of sight velocity map of the molecular gas traced by CO(3--2) observed using ALMA (\textit{left}) and the ionized gas traced by [OII] observed using MUSE (\textit{right}) in HE 0515-4414 Galaxy 1.\label{Fig:VelMaps}}
\end{figure*}


\bsp	
\label{lastpage}
\end{document}